\documentclass[conference]{IEEEtran}
\IEEEoverridecommandlockouts
\usepackage{cite}
\usepackage{amsmath,amssymb,amsfonts}
\usepackage{algorithmic}
\usepackage{graphicx}
\usepackage{textcomp}
\usepackage{xcolor}
\usepackage{enumitem}

\usepackage{mathtools}
\usepackage{algorithm}
\usepackage{array}
\usepackage[labelfont=rm,labelformat=parens]{subfig}
\usepackage{stfloats}
\usepackage{url}
\usepackage{verbatim}
\usepackage{bm}
\usepackage{booktabs}
\usepackage[T1]{fontenc}
\usepackage[utf8]{inputenc}
\def\BibTeX{{\rm B\kern-.05em{\sc i\kern-.025em b}\kern-.08em
    T\kern-.1667em\lower.7ex\hbox{E}\kern-.125emX}}
\begin{document}

\title{Ground Reflection-Aided TomoSAR Imaging \\ with 5G NR Signals \\

\author{Qiuyuan~Yang, Cunhua~Pan, Hong~Ren, Jiangzhou~Wang,~\IEEEmembership{Fellow,~IEEE}}

\thanks{Qiuyuan~Yang, Cunhua~Pan, Hong~Ren and Jiangzhou~Wang are with National Mobile Communications Research Laboratory, Southeast University, Nanjing 210096, China (email: 220241246, cpan, hren, j.z.wang@seu.edu.cn).}
}

\maketitle

\begin{abstract}
Tomographic synthetic aperture radar (TomoSAR) enables three-dimensional imaging by resolving targets along the elevation dimension, which is essential for environment reconstruction and infrastructure monitoring. A critical challenge in TomoSAR is the severe multipath propagation that causes ghost targets, range offsets, and elevation ambiguities. To address this, this paper proposes an enhanced Newtonized orthogonal matching pursuit (NOMP) algorithm to extract the delay, Doppler, and complex amplitude parameters of each propagation path, effectively separating line-of-sight (LoS) and multipath components prior to TomoSAR processing. Additionally, a height fusion strategy combining TomoSAR estimates with LoS-ground reflection delay-based inversion improves elevation accuracy. Simulation results demonstrate that the proposed method achieves improved positioning and elevation accuracy while effectively suppressing multipath-induced artifacts.
\end{abstract}

\begin{IEEEkeywords}
Integrated sensing and communication (ISAC), TomoSAR imaging, NOMP, ground reflection
\end{IEEEkeywords}

\section{Introduction}
With the vision of sixth-generation (6G) wireless networks, integrated sensing and communication (ISAC) has emerged as a fundamental technology to enable environment-aware intelligent systems\cite{liu2020joint}\cite{wang2021symbiotic}. By sharing spectrum, hardware platforms, and signal processing chains, ISAC tightly integrates conventional communication and radar sensing functionalities, significantly improving spectral efficiency while reducing system complexity and deployment cost. This paradigm is expected to play a crucial role in a wide range of applications, including low-altitude economy, autonomous transportation, smart cities, and industrial internet of things (IoT)\cite{zhu2024enabling}.

Meanwhile, 3D perception of complex environments has become increasingly important. Synthetic aperture radar (SAR) has been widely adopted due to its all-weather, all-day imaging capability and high spatial resolution\cite{koo2012new}. However, conventional SAR imaging is inherently limited to 2D  representations. Targets with different heights may be projected onto the same range–azimuth resolution cell, resulting in layover and height ambiguity. Tomographic SAR (TomoSAR) overcomes this limitation by exploiting multiple spatial baselines to resolve the elevation dimension, enabling true 3D reconstruction of distributed scatterers. Compared with interferometric SAR (InSAR), which can only retrieve a single dominant height per resolution cell, TomoSAR is capable of reconstructing volumetric scattering structures, making it particularly suitable for urban scenes and complex environments\cite{zhang2024target}\cite{tang2025cooperative}.

In recent years, significant research efforts have been devoted to TomoSAR imaging. \cite{zhang2022building} analysed triple-bounce scattering phenomena and proposed a multiple bounce scattering model to suppress ghost interference. The work \cite{guan20213} demonstrated 3D radar imaging using millimeter-wave signals, achieving angular and range resolution suitable for indoor scenarios. In \cite{10083170}, the 5G signal was exploited for passive radar processing, proving the feasibility of reference signals for target detection. In \cite{guo2024urban}, the geometric Z-structure constraint of buildings was exploited to improve the reconstruction quality of urban TomoSAR.

The main contributions of this paper are summarized as follows:

\begin{enumerate}[label=\arabic*)]
	\item We develop a TomoSAR imaging system utilizing 5G NR 
	synchronization signal block (SSB) as the reference signal. The proposed framework integrates communication and sensing functionalities. 
	\item We propose a height fusion strategy that integrates TomoSAR elevation estimates with height information inverted from the LoS-ground reflection delay difference, reducing elevation errors in multipath environments.
	\item We validate the effectiveness of the proposed cooperative ISAC framework and sensing scheme based on the simulation results.
\end{enumerate}

\section{SYSTEM MODEL}

\begin{figure}[!t]
	\centering
	\includegraphics[width=\columnwidth]{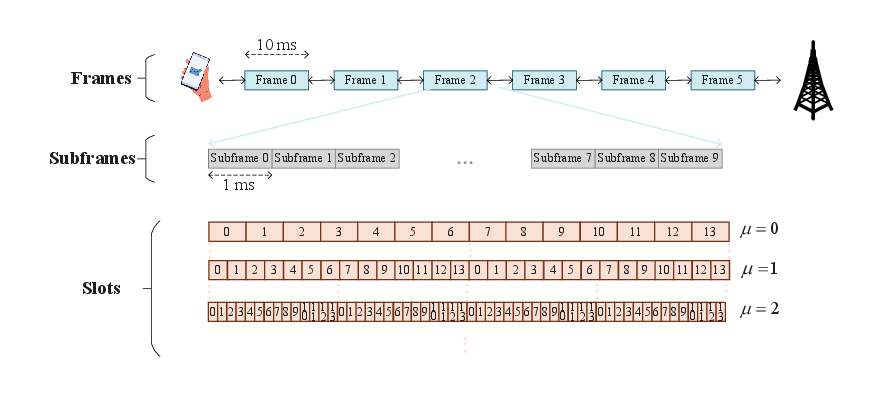}
	\caption{Frame structure in 5G NR.}
	\label{fig:fig2}
\end{figure}

\subsection{Wireless Frame Structure}
As illustrated in Fig.~\ref{fig:fig2}, the number of slots per subframe increases with the sampling interval of subcarrier spacing (SCS) configuration $\mu$. In this paper, the 
SSB is adopted as the reference 
signal for TomoSAR imaging. For SCS $\Delta f^{\mu}=2^{\mu}\times 15\left[ \mathrm{KHz} \right] , \mu \in \left\{ 0,1,...,6 \right\}$, the SSB bandwidth is $B_{\mathrm{SSB}}=240 \times \Delta f$. With $\Delta f = 120\left[ \mathrm{KHz} \right]$, the bandwidth is $28.8\mathrm{MHz}$, corresponding to a range resolution of: $\rho _r=\frac{c}{2B_{\mathrm{SSB}}}\approx 2.6\mathrm{m}$, where $c$ is the speed of light. The lower boundary value of the  numerology index $\mu$ can be directly determined as
\begin{equation}
	\mu \ge \left\lceil \log_2\left(\frac{c}{480 \times 15 \times 10^3 \times \rho_r} \right) \right\rceil,
\end{equation}
where $\lceil \cdot \rceil$ denotes the ceiling operator.

The maximum unambiguous range $R_{\mathrm{unamb}} = \frac{c}{2\Delta f}$ is constrained by SCS to avoid range aliasing in the radar. The upper boundary value of the  numerology index $\mu$ can be directly determined as
\begin{equation}
	\mu \le \left\lfloor \log_2\left(\frac{c}{2 \times 15 \times 10^3 \times R_{\mathrm{unamb}}} \right) \right\rfloor,
\end{equation}
where $\lfloor \cdot \rfloor$ denotes the flooring operator.

The SSB burst periodicity $T_{\mathrm{SSB}}$ determines the effective pulse repetition interval (PRI) for TomoSAR
\begin{equation}
    \mathrm{PRI}=T_{\mathrm{SSB}}=n_0 \cdot \frac{L_0}{10\times 2^\mu},
\end{equation}
where $n_0$ is the minimum SSB burst period in time slots, $L_0$ is the length of each wireless frame. The probe signals of TomoSAR need to be transmitted at a restricted Pulse repetition frequency (PRF) to prevent range and azimuth ambiguity
$\mathrm{PRF}=\frac{1}{\mathrm{PRI}} \geqslant 2\cdot \frac{2v_{\mathrm{unamb}}}{\lambda}=\frac{4f_cv_{\mathrm{unamb}}}{c}$, where the maximum unambiguous velocity $v_{unamb}=\frac{\lambda}{4T_0}=\frac{c\cdot \Delta f}{4f_c\left( 1+\Delta f\cdot T_{CP} \right)}$ is determined by the OFDM symbol duration $T_0$, $f_c$ is the carrier frequency, $T_{CP}$ is the time duration of the CP. The lower boundary value of the  numerology index $\mu$ can be directly determined as
\begin{equation}
	\mu \ge \left\lceil \log_2\left(\frac{4 f_c v_{\mathrm{unamb}} \times n_0 L_0}{10 \times c} \right) \right\rceil.
\end{equation}

Combining the above constraints, the numerology index $\mu$ for integrated communication and TomoSAR must satisfy:
\begin{equation}
	\begin{aligned}
		\mu_{\min} 
		&= \max \Biggl\{
		\begin{aligned}
			&\left\lceil \log_2\left( \frac{c}{480 \times 15 \times 10^3 \times \rho_r} \right) \right\rceil, \\
			&\left\lceil \log_2\left( \frac{4 f_c v_{\mathrm{unamb}} \times n_0 L_0}{10 \times c} \right) \right\rceil
		\end{aligned}
		\Biggr\}, \\
		\mu_{\max} 
		&= \left\lfloor \log_2\left( \frac{c}{2 \times 15 \times 10^3 \times R_{\mathrm{unamb}}} \right) \right\rfloor.
	\end{aligned}
\end{equation}

Thus, the proposed wireless frame structure is flexibly adjustable to ensure the efficient PRF management for TomoSAR imaging under different UAV configurations.

\begin{figure}[!t]
	\centering
	\includegraphics[width=\columnwidth]{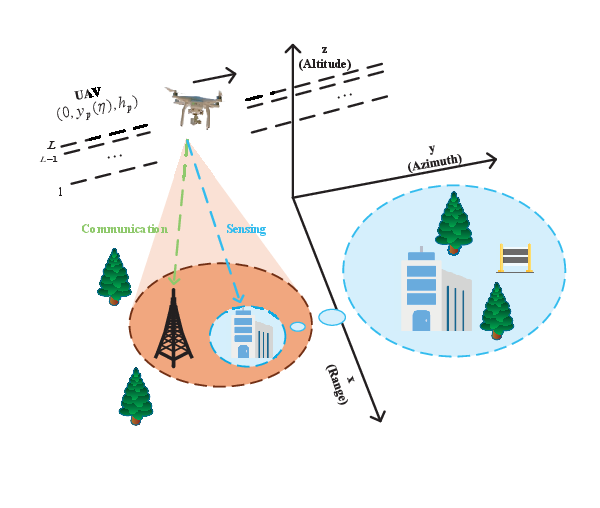}
	\caption{Monostatic OFDM TomoSAR geometry.}
	\label{fig:fig1}
\end{figure}

\subsection{TomoSAR Receiving Signal Model}
As shown in Fig.~\ref{fig:fig1}, we consider a monostatic TomoSAR platform carried by a UAV. The unmanned aerial vehicle (UAV) periodically transmits OFDM signals while receiving backscattered echoes from ground targets during flight. Without loss of generality, we assume that the UAV flies along the y-axis. The platform's motion enables azimuth-axis synthetic aperture, thereby achieving high-resolution imaging capability. 

Consider an OFDM transmitter \cite{zhu2009chunk} with $N$ subcarriers, with the subcarrier spacing and bandwidth being given as $\Delta f$ and $B = N\Delta f$. Then, the baseband time-domain OFDM signal within a symbol duration can be represented as
\begin{equation}
	s(t) = \frac{1}{\sqrt{N}} \sum_{k=0}^{N-1} S_k \exp\{j2\pi k \Delta f (t - T_{\mathrm{CP}})\}, \ t \in [0, T_0 + T_{\mathrm{CP}}],
\end{equation}
where $S_k$ satisfying $\sum_{k=0}^{N-1} |S_k|^2 = N$ denotes the modulation symbol on the $k$th subcarrier, which ensures that the average symbol power is one.

The instantaneous position of the UAV during the $i$th baseline flight can be expressed as $(0, y_p(\eta), h_i)$, where $\eta$ denotes the slow time index in SAR imaging, and $h_i = h_0 + i \cdot \Delta h$ represents the altitude of the $i$th flight with $h_0$ being the reference altitude and $\Delta h$ denoting the baseline spacing. The baseline index $i \in \{0, 1, ..., L-1\}$, where $L$ is the total number of baselines.

For stationary ground targets, when the UAV operates in swath mode and illuminates the target scene, the echo signal from the $m$th range cell can be expressed as
\begin{equation}
	r_m^{(i)}(t, \eta) = \sum_{p=0}^{P-1} \alpha_p^{(i)}(\eta) 
	s\left(t - \tau_{p}^{(i)}(\eta)\right) 
	e^{j2\pi f_{p}^{(i)}(\eta) t} + w_m(t, \eta),
\end{equation}
where $P$ denotes the number of multipath components, $\alpha _p\left( \eta \right)$ represents the complex gain of the $p$th path incorporating amplitude and phase, $\tau _{p}(\eta)$ denotes the propagation delay, $f_{p}\left( \eta \right)$ is the Doppler shift, $\eta$ is the slow-time index, and $w_m\left( t,\eta \right)$ represents additive noise. The path index $p=0$ corresponds to the LoS component when available, while $p \geqslant 1$ represents non-line-of-sight (NLoS) paths arising from single or multiple reflections.

The LoS path represents direct propagation from the transmitter to the target and back to the receiver. The time-variant delay of LoS path can be calculated as
\begin{equation}
    \tau _0\left( \eta \right) = \tau _{LoS}\left( \eta \right)= \frac{2R(\eta)}{c},
\end{equation}
where $R(\eta)$ is the slant range from the UAV to the target.

The power gain of the LoS path is given by
\begin{equation}
    \alpha_0(\eta) = \frac{\lambda \sqrt{G_{\mathrm{TX}} G_{\mathrm{RX}} 
    		\sigma_{\mathrm{RCS}}}}{(4\pi)^{3/2} R^2(\eta)} \cdot 
    e^{j\phi_0(\eta)},
\end{equation}
where $G_{\mathrm{TX}}$ and $G_{\mathrm{RX}}$ are the transmitter and receiver antenna gains, $\sigma _{\mathrm{RCS}}$ is the radar cross section of the targets, and $\phi_0(\eta) = -\frac{4\pi}{\lambda}R(\eta)$ is the phase shift.

The NLoS paths arise from reflections off environmental scatterers such as ground surfaces or building facades. The propagation delay of the NLoS path is given by
\begin{equation}
    \tau_p(\eta) = 
    \frac{R^{\mathrm{T,S}}_p(\eta) + R^{\mathrm{S,T}}_p(\eta) + 
    	2R^{\mathrm{S,O}}_p(\eta)}{c},
\end{equation}
where $R^{\mathrm{T,S}}_p(\eta)$ denotes the UAV-to-scatterer distance, $R^{\mathrm{S,O}}_p(\eta)$ denotes the scatterer-to-target distance. The complex gain of the ground reflection path is expressed as

\begin{equation}
    \begin{split}
    	\alpha_p(\eta) 
    	&= \Gamma_p \sqrt{G_{\mathrm{TX}} G_{\mathrm{RX}}} \cdot 
    	\frac{\lambda}{4\pi R^{\mathrm{T,S}}_p(\eta)} \\
    	&\quad \cdot \frac{\sqrt{\sigma_{\mathrm{RCS}}}}
    	{4\pi R^{\mathrm{S,T}}_p(\eta)} \cdot e^{j\phi_p(\eta)},
    \end{split}
\end{equation}
where $\Gamma_p$ is the reflection coefficient determined by the material properties of the $p$th scatterer, and 
$\phi_p(\eta) = -\frac{2\pi}{\lambda}(R^{\mathrm{T,S}}_p(\eta) + R^{\mathrm{S,T}}_p(\eta))$ is the corresponding phase shift.

In the presence of moving targets, Doppler-delay coupling introduces cross-terms in the 2D delay-Doppler spectrum 
$\mathbf{C}$, which may lead to false path detection.

\section{Algorithm Design}
To address the multipath-induced challenges, this section proposes a Multipath-Aware NOMP algorithm for high-precision 
3D localization, which consists of three stages: 
coarse estimation, Newton-based refinement, and height 
inversion using ground reflection paths\cite{11014534}. 

\subsection{Coarse Estimation}
Based on the received signal model in Section II, the channel observation vector after OFDM demodulation can be expressed as
\begin{equation}
	\mathbf{h}_s=\sum_{p=0}^{P-1}{\beta _p\left( \eta \right) \cdot \mathbf{a}\left( \tau _p, v_p \right) +\mathbf{w}},
\end{equation}
where $\beta _p\left( \eta \right) = \alpha _p(\eta) \cdot \exp^{-j2\pi f_c \tau_p(\eta)}$ is the equivalent complex path gain, and $\mathbf{a}\left( \tau _p, v_p \right) \in \mathbb{C} ^{\left| \varOmega _s \right|\times 1}$ is the steering vector with elements
\begin{equation}
    \left[ \mathbf{a}\left( \tau ,v \right) \right] _{\left( k,m \right)} = e^{-j2\pi k\varDelta f\tau}\cdot e^{j2\pi mT_0v},\left( k,m \right) \in \varOmega _s.
\end{equation}

The coarse estimation stage detects multipath components using FFT-accelerated correlation. Initialize the residual as $\mathbf{h}_r = \mathbf{h}_s$ and the path set as $\mathcal{P} =\varnothing$. The algorithm iteratively detects paths until the residual energy falls below threshold $\delta$. At each iteration, construct the residual matrix $\mathbf{R} \in \mathbb{C} ^{N\times M}$ and compute the 2D delay-Doppler correlation spectrum
\begin{equation}
    \mathbf{C} =\mathcal{F} _M\left\{ \mathcal{F} _{N}^{-1}\left\{ \mathbf{R} \right\} \right\}^{\mathrm{T}},
\end{equation}
The coarse parameter estimates are obtained by peak detection $\left( \hat{p}, \hat{q} \right)=arg\underset{p,q}{\max}|\left[ \mathbf{C} \right] _{p,q}|^2$, $\hat{\tau}_{coarse} = \frac{\hat{p}}{N\Delta f}$, $\hat{v}_{coarse} = \frac{\hat{q}-M/2}{MT_0}$, with amplitude $\hat{\beta}_{coarse} = \mathbf{a}^H\mathbf{h}_r/||\mathbf{a}||^2_2$. The path is added to $\mathcal{P}$, and the residual is updated as 
\begin{equation}
	\mathbf{h}_r\gets \mathbf{h}_r - \hat{\beta}_{coarse}\mathbf{a}\left( \hat{\tau}_{coarse},\hat{v}_{coarse} \right). 
\end{equation}
This process repeats until $\max|\left[ \mathbf{C}\right]_{p,q}|^2 <\delta$, yielding a set of $L$ detected paths $\mathcal{P}=\left\{ \left( \hat{\tau}_p,\hat{v}_p,\hat{\beta}_p \right) \right\} _{p=0}^{P-1}$.

\subsection{Newton-based Refinement}
The fine estimation stage refines parameters to off-grid positions using Newton iteration. For each detected path with parameters $\boldsymbol{\theta} = [\tau, v]^{\mathrm{T}}$, the refinement maximizes the log-likelihood function
\begin{equation}
	S(\boldsymbol{\theta}, \beta) = 2\Re\{\mathbf{h}_r^H 
	\mathbf{a}(\boldsymbol{\theta}) \beta\} - |\beta|^2 
	\|\mathbf{a}(\boldsymbol{\theta})\|_2^2.
\end{equation}
For a fixed $\boldsymbol{\theta}$, the optimal amplitude is given by
\begin{equation}
	\hat{\beta} = \frac{\mathbf{a}^H(\boldsymbol{\theta}) \mathbf{h}_r}
	{\|\mathbf{a}(\boldsymbol{\theta})\|_2^2}.
\end{equation}
Substituting $\hat{\beta}$ into $S(\boldsymbol{\theta}, \beta)$, 
the concentrated objective function becomes
\begin{equation}
	S(\boldsymbol{\theta}) = \frac{|\mathbf{a}^H(\boldsymbol{\theta}) 
		\mathbf{h}_r|^2}{\|\mathbf{a}(\boldsymbol{\theta})\|_2^2}.
\end{equation}

The Newton update rule is
\begin{equation}
	\boldsymbol{\theta}^{(i+1)} = \boldsymbol{\theta}^{(i)} 
	- \left[\ddot{S}\!\left(\boldsymbol{\theta}^{(i)}\right)
	\right]^{-1} \dot{S}\!\left(\boldsymbol{\theta}^{(i)}\right),
\end{equation}
where $\dot{S}(\boldsymbol{\theta})$ and $\ddot{S}(\boldsymbol{\theta})$ 
denote the gradient and Hessian of $S(\boldsymbol{\theta})$, respectively. The gradient is computed as
\begin{equation}
	\dot{S}(\boldsymbol{\theta}) = \frac{2}
	{\|\mathbf{a}(\boldsymbol{\theta})\|_2^2} \Re\left\{ 
	\mathbf{h}_r^H \dot{\mathbf{a}}(\boldsymbol{\theta}) 
	\hat{\beta}^* \right\},
\end{equation}
where $\dot{\mathbf{a}}(\boldsymbol{\theta}) = 
[\frac{\partial \mathbf{a}}{\partial \tau}, 
\frac{\partial \mathbf{a}}{\partial v}]$ is the Jacobian matrix 
with elements. The single-path Newton iteration continues until 
$\|\boldsymbol{\theta}^{(i+1)} - \boldsymbol{\theta}^{(i)}\| < \epsilon$ or a maximum number of iterations is reached.

{\color{red}
The global refinement iterates over all 
$\hat{P}$ detected paths in a cyclic manner 
for $N_{\mathrm{global}} = 10$ rounds. The 
convergence of this stage is guaranteed by 
the Refinement Acceptance Condition (RAC): each Newton 
step is accepted when it strictly 
improves the concentrated objective function 
$S(\hat{\boldsymbol{\theta}}_p)$, i.e.,
\begin{equation}
	S(\hat{\boldsymbol{\theta}}_p^{\text{new}}) 
	> S(\hat{\boldsymbol{\theta}}_p^{\text{old}}),
	\label{eq:RAC}
\end{equation}
which is equivalent to requiring
\begin{equation}
	\|\mathbf{h}_r^{(k+1)}\|^2 \leq 
	\|\mathbf{h}_r^{(k)}\|^2.
	\label{eq:nonincreasing}
\end{equation}
Therefore, the residual energy sequence 
$\{E_k\}$, where $E_k = 
\|\mathbf{h}_r^{(k)}\|^2$, is non-increasing 
and bounded below by zero\cite{mamandipoor2016newtonized}. By the Monotone 
Convergence Theorem, $\{E_k\}$ converges, 
and since each accepted step strictly 
decreases $E_k$, the parameter updates 
satisfy
\begin{equation}
	\|\hat{\boldsymbol{\theta}}_p^{(k+1)} - 
	\hat{\boldsymbol{\theta}}_p^{(k)}\| \to 0, 
	\quad k \to \infty,
	\label{eq:param_converge}
\end{equation}
confirming convergence to a stationary point 
of the joint objective. This convergence 
behavior is analogous to Theorem~1 of 
\cite{mamandipoor2016newtonized} for the 
outer detection loop of NOMP.

Furthermore, following the analytical 
framework of Theorem~3 and Lemma~1 in 
\cite{mamandipoor2016newtonized}, the 
residual energy after $n$ global rounds 
satisfies
\begin{equation}
	E_n \leq \frac{\eta^2}{n+1},
	\label{eq:rate}
\end{equation}
where
\begin{equation}
	\eta = \|\mathbf{h}_s\|_{\mathcal{A}} 
	\cdot \left(1 - \frac{2\pi}{\gamma}
	\right)^{-1},
	\label{eq:eta}
\end{equation}
where $\|\mathbf{h}_s\|_{\mathcal{A}}$ 
denotes the atomic norm of $\mathbf{h}_s$ 
with respect to the steering vector 
dictionary $\{\mathbf{a}(\tau,v)\}$, and 
$\gamma$ is the oversampling factor. 
Equation~\eqref{eq:rate} establishes an 
$\mathcal{O}(1/n)$ decay rate for the 
residual energy. In our scenario, 
\eqref{eq:rate} guarantees that after 
$N_{\mathrm{global}} = 10$ rounds, the 
residual energy satisfies
\begin{equation}
	E_{10} \leq \frac{\eta^2}{11},
	\label{eq:E10}
\end{equation}
i.e., reduced to at most $\eta^2/11$ of 
its initial value. Furthermore, 
\cite{mamandipoor2016newtonized} 
demonstrates empirically that cyclic Newton 
refinement exhibits diminishing returns 
beyond $3$--$5$ rounds even for $K = 16$ 
components (Fig.~12 in 
\cite{mamandipoor2016newtonized}), 
confirming that $N_{\mathrm{global}} = 10$ 
provides a sufficient and conservative 
convergence margin for the global 
refinement stage of 
ED-NOMP.
}

In each round, for the $p$th path, the residual is first reconstructed by adding back its contribution:
\begin{equation}
	\mathbf{h}_r^{(p)} = \mathbf{h}_r + \hat{\beta}_p 
	\mathbf{a}(\hat{\boldsymbol{\theta}}_p).
\end{equation}
Then the Newton refinement is applied to update the parameters 
of the $p$th path using $\mathbf{h}_r^{(p)}$ as the observation. 
The updated parameters are obtained as
\begin{equation}
	\hat{\boldsymbol{\theta}}_p^{\mathrm{new}} = 
	\hat{\boldsymbol{\theta}}_p - 
	\left[\ddot{S}(\hat{\boldsymbol{\theta}}_p)\right]^{-1} 
	\dot{S}(\hat{\boldsymbol{\theta}}_p),
\end{equation}
and the amplitude is updated accordingly:
\begin{equation}
	\hat{\beta}_p^{\mathrm{new}} = \frac{\mathbf{a}^H
		(\hat{\boldsymbol{\theta}}_p^{\mathrm{new}}) \mathbf{h}_r^{(p)}}
	{\|\mathbf{a}(\hat{\boldsymbol{\theta}}_p^{\mathrm{new}})\|_2^2}.
\end{equation}
After updating the $p$th path, the global residual is updated as
\begin{equation}
	\mathbf{h}_r \gets \mathbf{h}_r^{(p)} - \hat{\beta}_p^{\mathrm{new}} 
	\mathbf{a}(\hat{\boldsymbol{\theta}}_p^{\mathrm{new}}).
\end{equation}

This cyclic refinement process is repeated for all $\hat{P}$ paths within each round.

\subsection{Ground reflection-based inversion}
The LoS path corresponds to the minimum propagation delay among all detected paths: $\hat{\tau}_{\mathrm{LoS}} = \min_{p \in \mathcal{P}}\hat{\tau}_p$. The target range and radial velocity are estimated as
\begin{equation}
	\hat{R} = \frac{c \hat{\tau}_{\mathrm{LoS}}}{2}, \quad 
	\hat{v}_r = \frac{\lambda \hat{v}_{\mathrm{LoS}}}{2}.
\end{equation}

For the monostatic configuration, the LoS and ground reflection 
path lengths are $R_{\text{LoS}} = \sqrt{R_g^2 + (h_i - z_t)^2}$ 
and $R_{\text{GR}} = \sqrt{R_g^2 + (h_i + z_t)^2}$ respectively, where $R_g$ is the horizontal distance from the platform, $z_t$ is the target height.

Under the far-field assumption where $R_g \gg h_i, z_t$, the path length difference is then
\begin{equation}
	R_{\mathrm{GR}} - R_{\mathrm{LoS}} \approx 
	\frac{(h_i + z_t)^2 - (h_i - z_t)^2}{2R_g} = \frac{2 h_i z_t}{R_g}.
\end{equation}
Since the slant range $R \approx R_g$ , the delay difference can be expressed as
\begin{equation}
	\Delta \tau_{\mathrm{GR}} = \frac{2(R_{\mathrm{GR}} - R_{\mathrm{LoS}})}{c} 
	\approx \frac{4 h_i z_t}{c R}.
\end{equation}

For each detected path $p \neq \mathrm{LoS}$, the delay difference $\Delta \hat{\tau}_p = \hat{\tau}_p - \hat{\tau}_{\mathrm{LoS}}$ implies a candidate target height:
\begin{equation}
	\hat{z}_t^{(p)} = \frac{c \cdot \hat{R} \cdot \Delta \hat{\tau}_p}{4 h_i}.
\end{equation}

A path is identified as a ground reflection candidate if the implied height falls within a physically reasonable range, i.e., $\hat{z}_t^{(p)} \in [z_{\min}, z_{\max}]$, where $z_{\min}$ and $z_{\max}$ define the expected target height bounds.

Additionally, ground reflection experiences attenuation due to the reflection coefficient and longer propagation path. The amplitude ratio between a candidate path and the LoS path should satisfy
\begin{equation}
	\rho_p = \frac{|\hat{\beta}_p|}{|\hat{\beta}_{\mathrm{LoS}}|} 
	\in [\rho_{\min}, \rho_{\max}],
\end{equation}
where $\rho_{\min}$ and $\rho_{\max}$ correspond to typical ground reflection coefficients determined by material properties.

For TomoSAR imaging with $L$ baseline positions, the LoS path gains $\{\hat{\beta}^{(i)}_{\mathrm{LoS}}\}_{i=0}^{L-1}$ across different baselines encode elevation information. The TomoSAR-based height estimate is obtained
\begin{equation}
	\hat{z}_A = \arg\max_{z} \;
	\frac{1}{\mathbf{a}^H(z)\,\mathbf{E}_n \mathbf{E}_n^H\,\mathbf{a}(z)},
\end{equation}
where $\mathbf{E}_n$ contains the noise subspace eigenvectors derived from the covariance matrix of the baseline observations. The ground reflection path provides a height estimate:
\begin{equation}
	\hat{z}_{\mathrm{B}} = \frac{c \cdot \hat{R} \cdot 
		\Delta \hat{\tau}_{\mathrm{GR}}}{4 h_i}.
\end{equation}
The final elevation estimate fuses both sources with confidence 
weighting:
\begin{equation}
	\hat{z} = \frac{w_{\mathrm{A}} \hat{z}_{\mathrm{A}} + 
		w_{\mathrm{B}} \hat{z}_{\mathrm{B}}}{w_{\mathrm{A}} + w_{\mathrm{B}}},
\end{equation}
where $w_{\mathrm{A}}$ and $w_{\mathrm{B}}$ reflect the estimation reliability based on SNR and geometric consistency. When no valid ground reflection path is detected, we set $w_{\mathrm{B}} = 0$ and rely solely on the TomoSAR estimate.

\subsection{Computational Complexity Analysis}
Table~\ref{tab:complexity} compares the computational complexity of different algorithms. The NOMP algorithm consists of three stages: coarse estimation with complexity $\mathcal{O}(LNM(\log N + \log M))$, local Newton refinement with $\mathcal{O}(L\hat{P} N_{\text{iter}}|\Omega_s|)$, and global refinement with $\mathcal{O}(LN_{\text{global}}|\Omega_s|\hat{P}^2 + L\hat{P}^3)$, where $N_{\text{iter}}$ denotes the number of Newton iterations per path. The proposed ED-NOMP introduces an additional complexity of $\mathcal{O}(L^3 + LN_z)$ for joint TomoSAR elevation estimation across $L$ baselines, where $N_z$ is the number of elevation grid points.

The computational cost of ED-NOMP remains feasible, as the dominant complexity term $\mathcal{O}(LN_{\text{global}}|\Omega_s|\hat{P}^2+L\hat{P}^3)$ scales modestly with the number of multipath components encountered in practice.

\begin{table}[t]
	\centering
	\caption{Computational Complexity Comparison}
	\label{tab:complexity}
	\begin{tabular}{|l|l|}
		\hline
		\textbf{Algorithm} & \textbf{Computational Complexity} \\
		\hline
		OMP & $\mathcal{O}(LNM(\log N+\log M)) + \mathcal{O}(L\hat{P}^3|\Omega_s|)$ \\
		\hline
		MUSIC & $\mathcal{O}(L^3 + LN_z)$ \\
		\hline
		NOMP & $\mathcal{O}(LNM(\log N+\log M)) + \mathcal{O}(L\hat{P} N_{\text{iter}}|\Omega_s|)$ \\
		& $+ \mathcal{O}(LN_{\text{global}}|\Omega_s|\hat{P}^2+L\hat{P}^3)$ \\
		\hline
		ED-NOMP & $\mathcal{O}(LNM(\log N+\log M)) + \mathcal{O}(L\hat{P} N_{\text{iter}}|\Omega_s|)$ \\
		& $+ \mathcal{O}(LN_{\text{global}}|\Omega_s|\hat{P}^2+L\hat{P}^3)$ \\
		& $+ \mathcal{O}(L^3 + LN_z)$ \\
		\hline
	\end{tabular}
	\vspace{1mm}
\end{table}

\begin{figure*}[htbp]
	\centering
	\subfloat[]{%
		\includegraphics[width=0.24\textwidth]{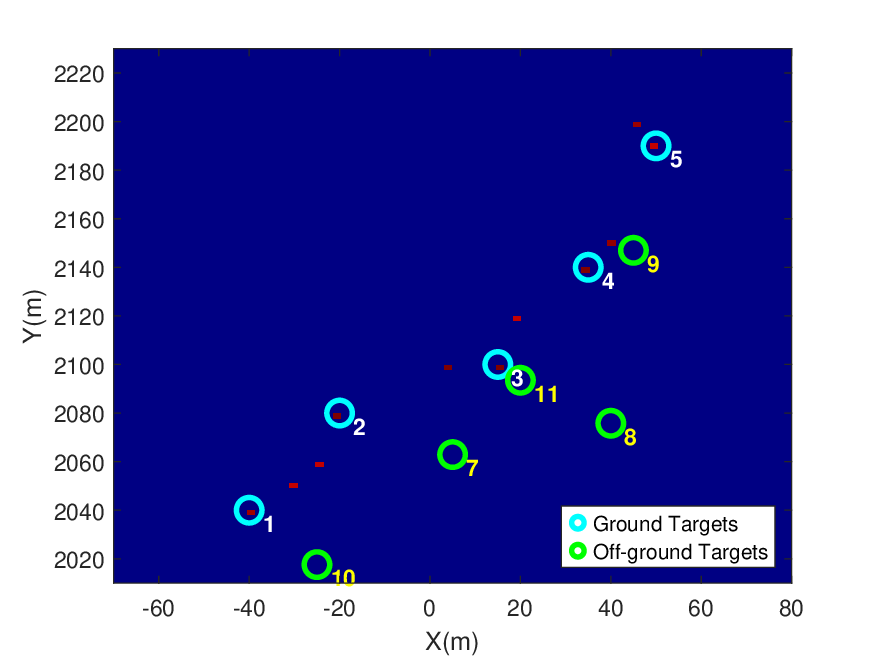}
		\label{fig:point_a}
	}
	\hfill
	\subfloat[]{%
		\includegraphics[width=0.24\textwidth]{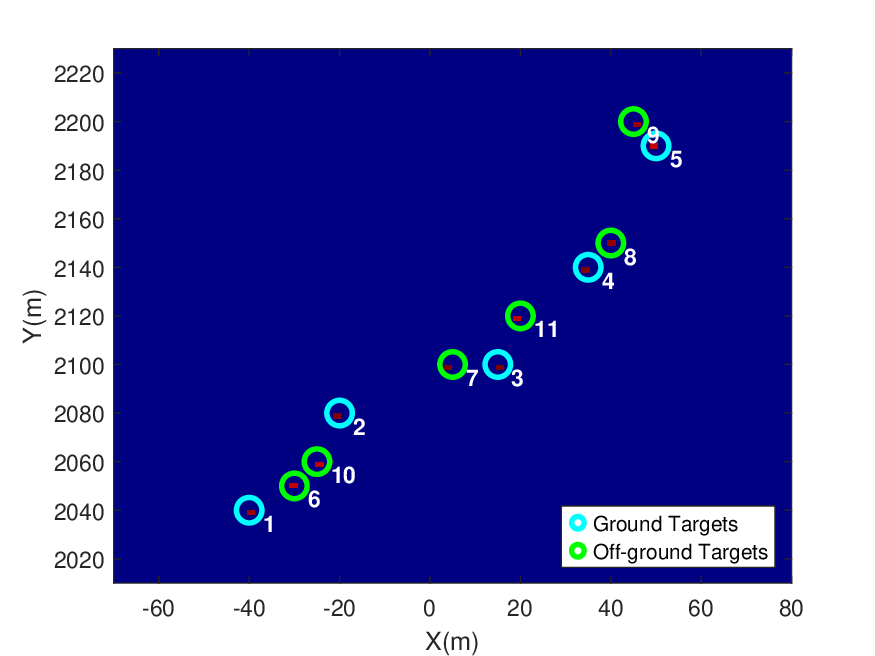}
		\label{fig:point_b}
	}
	\hfill
	\subfloat[]{%
		\includegraphics[width=0.24\textwidth]{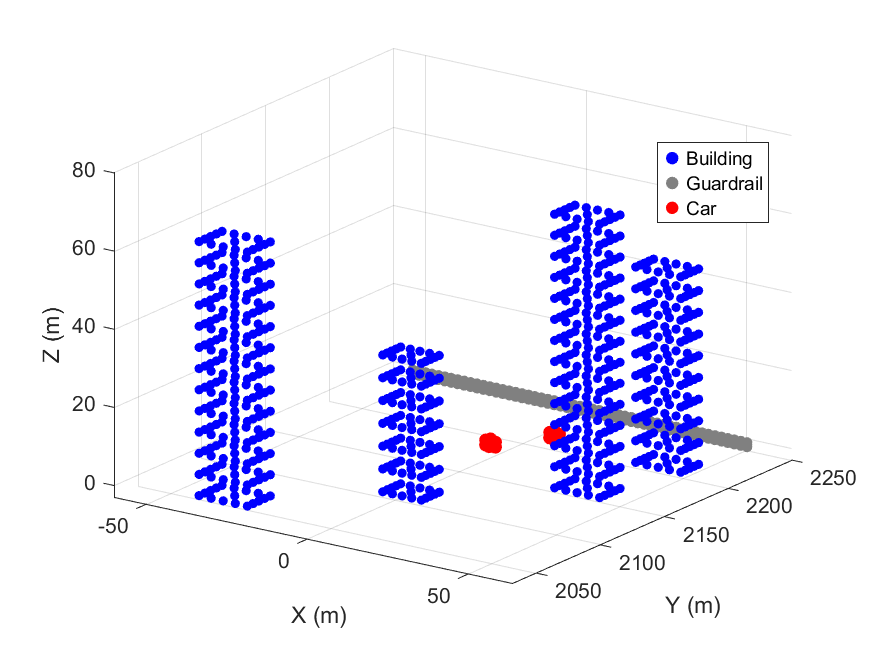}
		\label{fig:extend_a}
	}
	\hfill
	\subfloat[]{%
		\includegraphics[width=0.24\textwidth]{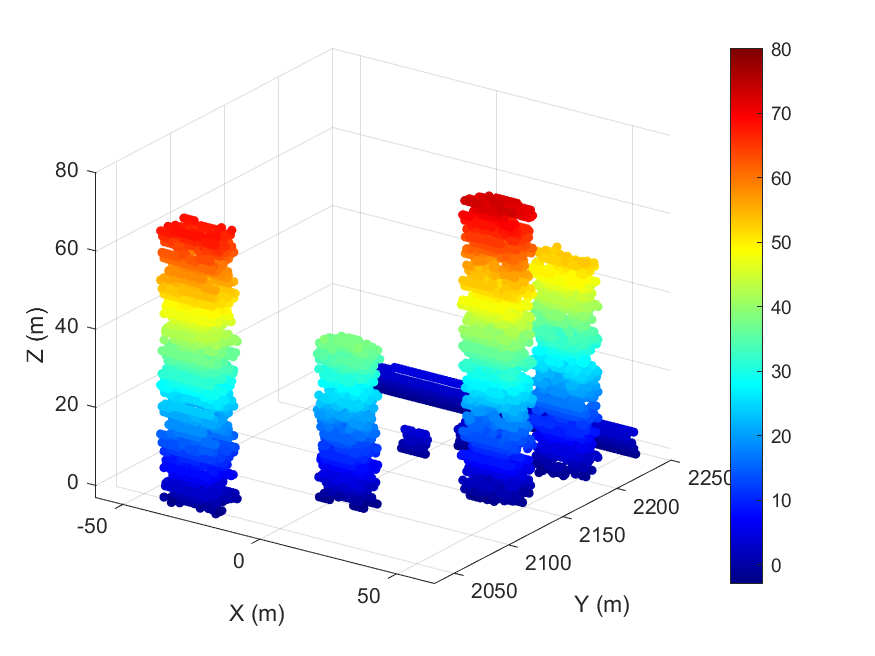}
		\label{fig:extend_b}
	}
	
	\caption{Imaging results. (a) SAR image with ground phase. (b) SAR image after ground phase removal. (c) True 3D target distribution. (d) Reconstruction result.}
	\label{fig:fig4}
\end{figure*}

\section{SIMULATION RESULTS}

This section evaluates the performance of the proposed algorithm through simulation experiments. The simulation parameters are shown in Table~\ref{tab:params}.

\begin{table}[htbp]
	\caption{Simulation parameters}
	\label{tab:params}
	\centering
	\begin{tabular}{cc}
		\toprule
		\textbf{Parameter} & \textbf{Value} \\
		\midrule
		Carrier frequency $f_c$ & 26 GHz \\
		Subcarrier spacing $SCS$ & 120 KHz \\
		Cyclic prefix length $M_{\mathrm{CP}}$ & 288 \\
		FFT size $M_{\mathrm{FFT}}$ & 4096 \\
		Bandwidth $B$ & 400 MHz \\
		Pulse repetition frequency $PRF$ & 800Hz \\
		Baseline spacing $\Delta h$ & 2m \\
		UAV reference altitude $h_0$ & 1 KM \\
		UAV velocity $v_p$ & 40 m/s \\
		
		\bottomrule
	\end{tabular}
\end{table}

Fig.~\ref{fig:fig4} presents the TomoSAR imaging results. Fig.~\ref{fig:fig4}(a) shows the SAR image with ground phase, where the target positions are distorted due to the flat-earth 
phase effect. After ground phase removal, Fig.~\ref{fig:fig4}(b) demonstrates improved target localization with reduced position offsets. Fig.~\ref{fig:fig4}(c) illustrates the true 3D target distribution, including four buildings, two stationary cars and a row of guardrails. The corresponding 3D reconstruction result in Fig.~\ref{fig:fig4}(d) shows that the proposed algorithm recovers the spatial structure of the targets.

Fig. 4 shows the RMSE performance comparison 
among ED-NOMP, NOMP and OMP algorithm with 
different numbers of baselines ($L= 6, 8, 10$). OMP shows the worst performance due to its significant inter-path interference unmitigated. The ED-NOMP($L$ = 6) outperforms NOMP($L$ = 6) , corresponding to an improvement of approximately 60\%. The results confirm that fusing ground reflection-based height inversion with TomoSAR effectively improves elevation estimation accuracy.

\begin{figure}[!t]
	\centering
	\includegraphics[width=\columnwidth]{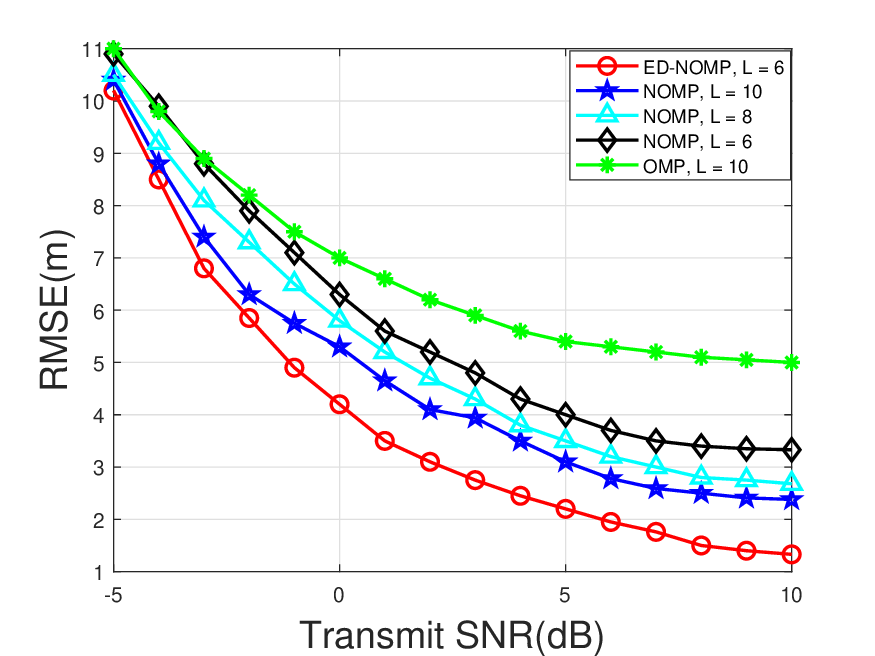}
	\caption{RMSE Performance Comparison.}
	\label{fig:fig5}
\end{figure}

\section{Conclusion}

In this paper, we proposed a novel UAV-TomoSAR framework for 3D reconstruction. The ED-NOMP algorithm effectively exploits ground-reflected multipath components to enhance elevation resolution. It is worth noting that practical hardware impairments such as phase noise and timing offset may affect system performance, and developing compensation methods remains an important direction for future work.

\bibliographystyle{IEEEtran}
\bibliography{reference}
\end{document}